\documentstyle[12pt]{article}
\begin{document}
\begin{titlepage}
\hspace{9 cm} ULB--PMIF--92/10

\vspace{5 cm}

\begin{centering}

{\huge Bianchi Cosmological Models and Gauge Symmetries}\\

\vspace{1 cm}

{\large Olivier Coussaert$^1$ and Marc Henneaux $^2$  \\
Facult{\'e} des Sciences, Universit{\'e} Libre de Bruxelles,\\
Campus Plaine C.P. 231, B-1050 Bruxelles, Belgium}\\
\end{centering}

\vskip 5 cm

\footnotesize{ $^1$ Chercheur IRSIA}

\footnotesize{ $^2$ Ma\^\i tre de Recherches au
Fonds National de la Recherche Scientifique. Also at Centros de
Estudios Cient\'\i ficos de Santiago, Casilla 16443,Santiago 9, Chile.}

\end{titlepage}

\begin{abstract}

We analyze carefully the problem of gauge  symmetries for Bianchi
models, from both the geometrical and dynamical points of view. Some of
the geometrical definitions of gauge symmetries (=``homogeneity
preserving diffeomorphisms'') given in the literature do not incorporate
the crucial feature that local gauge transformations should be independent
at each point of the manifold of the independent variables ( = time for
Bianchi models), i.e, should be arbitrarily localizable ( in time).
We give a geometrical definition of homogeneity preserving
diffeomorphisms that does not possess this shortcoming. The proposed
definition has the futher advantage of coinciding with the dynamical
definition based on the invariance of the action ( in Lagrangian or
Hamiltonian form). We explicitly verify the equivalence of the Lagrangian
covariant phase space with the Hamiltonian reduced phase space. Remarks
on the use of the Ashtekar variables in Bianchi models are also given.
\end{abstract}

\section{Introduction}

It is a well known fact that the distinction between global and local
symmetries is somewhat subtle in the case of
spatially homogeoneous cosmological models of the Bianchi type
 \cite{r1,r2,r3}.
There are indeed at least three different ways to define the gauge
symmetries  \footnote{We restrict the analysis of this paper to class A
Bianchi cosmologies as these are the only ones for which one can inject
the homogeneous form of the metric in the action prior to computing the
variationnal derivatives \cite{r4,r5,r6}}:
        \begin{itemize}
\item[(i)] geometrically, as those diffeomorphisms that preserve manifest
homogeneity;
        \item[(ii)] dynamically, from a spacetime covariant point of view,
 as those
transformations that leave the action invariant and that
are local in time; and
        \item[(iii)] dynamically, from the Hamiltonian point of view, as those
canonical transformations that are generated by the first class
constraints.
\end{itemize}

It is not a priori clear that these three different points of view should
yield identical conclusions. Futhermore, while the dynamical approaches
based on (ii) and (iii) yield unambiguous answers, the geometric
approach is
not so well defined since it depends on what one means exactly by
``diffeomorphisms preserving manifest homogneity''.

The purpose of this paper is to show that the three different concepts of
gauge symmetries are actually the same provided one adopts a natural
definition of ``homogeneity preserving diffeomorphisms''. This definition
agrees with the one given in \cite{r1,r2} but differs from that
 proposed in \cite{r3}. We
first consider the metric formulation of Bianchi models and then briefly
comment on the formulation in terms of Ashtekar variables.

\section{ Geometric definition of gauge symmetries}

\subsection{ First definition }

Bianchi cosmological models are spacetimes invariant under a three-
dimensional group of isometries G acting simply transitively on three -
dimensional spacelike hypersurfaces \cite{r7}.

In Bianchi models, one is thus given an action of $G$
 on the manifold $\cal M$ ,
 \begin{eqnarray}
G & \longrightarrow Diff({\cal M}) \\
\nonumber R \in  G &\longmapsto h_R \in Diff({\cal M})
\end{eqnarray}

and the spacetime metric $g$ fulfills
\begin{equation}
 h^*_R  g = g
\end{equation}

Since the action of G is part of the given geometric data,
it is natural to
define the subgroup $ H  \subset Diff({\cal M})   $
of ``homogeneity preserving
diffeomorphisms'' as the set of diffeomorphisms which preserve the
action of $G$, i.e.  which commute with each $h_R, R \in G          $
 \begin{equation}
\label{3}
  f \in H  \leftrightarrow f h_R = h_R f \hskip 2 cm  \forall R \in G
\end{equation}

If one performs a diffeomorphism $k$ that does not belong to $H$,
 one can
detect it by just comparing the action of $G$ before
 and after $k$ has been
performed. The group $H$ defined by (\ref{3}) is the centralizer
 of $G$ in $Diff( \cal M)$.

In order to determine the general solution of (\ref{3}) , we observe
 that the
manifold $\cal M$ is foliated by the three-dimensional
 hypersurfaces $\Sigma$
of transitivity of $G$. It is convenient
to introduce a non-vanishing vector
field $v$ transverse to these hypersurfaces and such that the one -
parameter group $\{ exp x^0 v \}$ commutes with $G$,
 \begin{equation}
\label{4}
( exp\  x^0 v ) h_R = h_R ( exp \ x^0 v )
\end{equation}

This can be done by picking an arbitrary curve $\gamma : x^0 \mapsto
\gamma  ( x^0 )      $ everywhere transverse to the hypersufaces $
\Sigma$. One then constructs the family of curves $x^0 \mapsto
h_R(\gamma (x_0) )$.  Since $G$ acts simply transitively,
there is one and only
one such curve through each point. One defines $v$
to the tangent vector field
to the family. The vector field $v$  fulfills (\ref{4}) and $x^0$
is constant along
the hypersurfaces of transitivity.

Let us now choose a basis in the Lie algebra $\cal G$ of $G$ and
 take the
corresponding Killing vectors $\xi_a (a=1,2,3) $ tangent to the
hypersurfaces $\Sigma$.
The vector fields $v$ and $\xi_a$ provide a basis of the tangent
 space at each point of
the spacetime manifold $\cal M$. One has

\begin{eqnarray}
\label{5}
& [ v , {\xi}_{a} ] = 0  \\
& [ \xi_{a} , \xi_{b} ] = C_{ab}^c  {\xi}_{c}  \nonumber
\end{eqnarray}
where  $ C_{ab}^c$ are the structure constants of $\cal{G}$.

Let $X_a$ be a basis of invariant vector fields fulfilling
also $ [v, X_a ] =
0$,
\begin{eqnarray}
\label{6}
& [ v , X_a ] = 0 \\
& [ \xi_a , X_b ] = 0 \nonumber
\end{eqnarray}
One may take $X_a = \xi_a $   on the curve given above.
In that case, on
finds in addition to (\ref{5})-(\ref{6})
\begin{equation}
 [X_a , X_b ] = - C_{ab}^c  X_c
\end{equation}

The most general ``infinitesimal'' \footnote{These are the only ones
considered here.} diffeomorphism commuting with $G$ is determined by a
vector field $\xi$ fulfilling
\begin{equation}
\label{8}
 [ \xi , \xi_a ] = 0
\end{equation}
If one expands $\xi $ in the basis $\{ v ,  X_a \}$,
\begin{equation}
\label{9}
\xi = \mu v + \mu^a X_a
\end{equation}
one gets
\begin{equation} \partial _{\xi^a} \mu = 0 , \hskip 2 cm
 \partial_{\xi^a} \mu^b = 0,
\end{equation}
 i.e.,
\begin{equation}
\label{11}
  \xi = \mu (x^0) v +  \mu^a ( x^0 ) X_a \end{equation}

Thus, a homogeneity preserving diffeomorphism  depends on four
functions of
$x^0$, irrespectively of the homogeneity group and of the spatial
topology. This is quite a satisfactory result, since in the
non-homogeneous
case, a general diffeomorphism depends also on four functions ( which
may be non-homogeneous). Futhermore, the fact  that  (\ref{11}) involves
arbitrary functions of $x^0$ implies that one can arbitrarily localize
the
homogeneity preserving diffeomorphisms in time ( there is, of course, no
hope to localize them in space in a spatially homogeneous context). This
localizability property appears to be an essential feature of gauge
(=``local'') transformations.

We close this subsection by observing that (\ref{11}) acts though
the adjoint
action on the invariant frame $\{ v , X_a \}$ ,
 \begin{eqnarray}
\label{12}
\delta v & = [ \xi , v ] =
- \frac{d\mu}{dx^0}v - \frac{d\mu^a}{dx^0} X_a \\
  \delta X_a & = [ \xi , X_a ] = - \mu^b C_{ab}^c X_c \nonumber
\end{eqnarray}

This yields
\begin{eqnarray}
\label{13}
\delta g_{00}  = \mu \frac{dg_{00}}{dx^0} + 2 \frac{dg_{00}}{dx^0}
   + 2 g_{0a} \frac{d\mu^a}{dx^0} \nonumber \\
\delta g_{0a}  = \mu \frac{dg_{0a}}{dx^0} + g_{0a} \frac{d\mu}{dx^0}
+ g_{ab} \frac{d\mu^b}{dx^0} + \mu^d C_{da}^b g_{0b} \\
\delta g_{ab}   = \mu \frac{dg_{ab}}{dx^0} + \mu^d ( C_{da}^c g_cb +
   C_{db}^c g_{ac} ) \nonumber
\end{eqnarray}
for the components $g_{\lambda \mu }(x^0)$ of the invariant metric
in the
invariant frame $\{ v , x_a \}$

The transformations (\ref{13}) are irreductible if and only if
the adjoint
representation is faithful, i.e., if the center of $\cal{G}$  vanishes.
 Otherwise,
there exist non trivial gauge parameters for which the transformations
(\ref{13}) are zero. These are given
by $ \mu = 0 , \frac{d\mu^a}{dx^0} =
0$ with $\mu^a   $ a non zero solution of $ \mu^a C_{ab}^c = 0$.
  An irreductible
description of the gauge transformations
 can be obtained as follows. Let $\xi_A$ be a basis of the
center of $\cal{G}$  $( C_{Ab}^d = 0 )$
and $\xi_\alpha$ supplementary vectors so
that $\{ \xi_A, \xi_\alpha \}$ is a basis of $\cal{G}$. On can replace
$\frac{d\mu^a}{dx^0}      $ by the gauge parameter $\epsilon ^A$ without
violating locality in time ( since $\mu^A$ does
 not appear in (\ref{13})), which
yields
\begin{eqnarray}
\label{14}
\delta g_{00}  = \mu \frac{dg_{00}}{dx^0} + 2 \frac{dg_{00}}{dx^0}
   + 2 g_{0A} \epsilon^A +
 2 g_{0\alpha} \frac{d\mu^\alpha}{dx^0} \nonumber \\
\delta g_{0a}  = \mu \frac{dg_{0a}}{dx^0} + g_{0a} \frac{d\mu}{dx^0}
+ g_{aB} \epsilon^B+ g_{a\beta} \frac{d\mu^\beta}{dx^0} + \mu^\delta
C_{\delta a}^b g_{0b} \\
\delta g_{ab}  = \mu \frac{dg_{ab}}{dx^0} + \mu^\delta
( C_{\delta a}^c g_{cb}
+ C_{\delta b}^c g_{ac} ) \nonumber
\end{eqnarray}

(For more information on reductible gauge transformations, see for
instance \cite{r8} chapter III).

\subsection{ Second definition}

A weaker form of ''homogeneity preserving diffeomorphisms'' has
 been
proposed in \cite{r3}. These are the diffeomorphisms which
map any spatially
homogeneous metric on a spatially homogeneous metric.
For this to be the
case, one must have in place of (\ref{8})
\begin{equation}
 \label{15}
 [ \xi , \xi_a ] = \mu_a^b \xi_c
\end{equation}
where the matrix $\mu_a^b$ is constant. Clearly,
any solutions of (\ref{8}) is
also a solution of (\ref{15}). The converse is not true for non compact
cosmologies of type $\overline{ \underline{I}}$,
 $\overline{ \underline{ II}}$, $\overline{ \underline{ VI}}_0$ or
$\overline{ \underline{VII}}_0$
 \cite{r3}. In finite term, the equation (\ref{15}) becomes
 $$ f h_R f^{-1} = h_{R'} $$
Thus, $f$ is a homogeneity preserving diffeomorphism according
to the second defintion if and only
if its belongs to the normalizer of $G$ in $Diff( \cal M)$.

Let us now determine which solutions of (\ref{15}) can be localized in
an
arbitrarily small spacetime interval. If $\xi(x^0)$ is zero for $ x^0
\not\in [ x^0_1 , x^0_2]$, one has $\mu_a^b = 0$ since
 $[\xi , \xi_a ] = 0$
 for $ x^0  \not\in [ x^0_1 , x^0_2] $ and  $ \mu_a^b  $
is constant. Hence, the only homogeneity preserving diffeomorphisms of
\cite{r3} that can differ from the identity in a finite time interval
are precisely
the homogeneity preserving diffeormorphisms of the previous subsection.
If one feels that the concept of ``gauge transformation''
( ``of the second
kind '' as one used to say in the old days) should involve in a crucial
manner the freedom to change the local reference frame independently at
each time ( localizability of the transformation), then, one must impose
the condition $\mu^a_b = 0$ in (\ref{15}) and one falls back on the first
definition given
previously. This point has not be addressed  in \cite{r3}.

\section{ Lagrangian definition of gauge symmetries}

\subsection{ Gauge transformations}

The  previous considerations are purely kinematical and do no refer
to the
action principle. To decide the issue of which geometrical definition is
preferable, one should introduce physical
considerations, namely, study the transformation properties of the action
principle.

The equations for the Bianchi models of class A can be derived from the
Hilbert action in which the symmetry of the metric is enforced before one
computes the variationnal derivatives (``reduced variationnal
principle'')\cite{r5,r6}
\begin{equation}
\label{16}
S[g_{ab}, N(x^0), N^a(x^0) ] = \int dx^0  L(g_{ab}, \frac{dg_{ab}}{dx^0}, N,
  N^a)
\end{equation}

The lagrangian contains reference to the homogeneity group through the
structure constants $C_{bc}^{a}$ that appear as parameters. It explicit
form is\cite{r5,r6}
\begin{equation}
\label{17}
  L(g_{ab}, \frac{dg_{ab}}{dx^0}, N, N^a) = \int dx^3 N \sqrt{g}
 (R - K^c_c
K^d_d + K^d_c K^c_d)
\end{equation}
where:
\begin{eqnarray}
 R= -\frac{1}{2} (2 {C^c}_{cd} {{C^a}_a}^d + C^{dca} C_{cda} +
\frac{1}{2} C^{cda} C_{cda} \nonumber \\
K_{ab} = \frac{1}{2N} [ ({C_{ab}}^c + {C_{ba}}^c ) N_c -
\frac{dg_{ab}}{dx^0}]
\nonumber
 \end{eqnarray}
with:
$$ C_{abc} = g_{ad} {C^d}_{bc}, {{C^a}_a}^c = g^{cd} {C^a}_{ad} $$
$$ C^{abc} = g^{bd} g^{ce} {C^a}_{de}$$
Note that the shift $N^a$ appears only through the combinaision
 $N^a C_{ab}^c$,
so that the components of $N^a$ in the center drop out.

A gauge transformation of (\ref{16}) is a transformation of the
dynamical
variables which must fulfills two conditions \cite{r8}:
\begin{itemize}

\item[(i)] it leaves the action invariant ( up to a boundary term); and

\item[(ii)] it is localizable; i.e., it must depend on arbitrary functions
ofthe
coordinates, which can be chosen so that the transformations reduce to
the identity outside a finite region. In the present case where we impose
spatial homogeneity, the independent coordinates reduce to $x^0$ and a
finite
region is typically a finite interval $[ x^0_1 , x^0_2 ]$.
 \end{itemize}
 It is
the second  point  that makes the distinction between gauge (``local'')
transformation and global symmetry transformations.
Gauge symmetries lead to non uniqueness of the Cauchy problem while
 global
symmetries do not.

It should be stressed that the definition (i)-(ii) of the gauge symmetries
isentirely ``covariant'' i.e.,  formulated from a spacetime point of view.
 The
fact that $x^0$ plays a priviledged role is
a result of the spatial homogeneity
alone and not not of the Hamiltonian splitting
of spacetime into space and time.

It is easy to verify that the transformations (\ref{13}) or (\ref{14})
when rewritten
in terms of $g_{ab}$ , the lapse and the shift, leave
the action (\ref{16}) invariant.
Furthermore, as we have discussed, they are local
in the sense of (ii). Thus,
they define gauge symmetries. They actually provide
a complete set of gauge
symmetries, because once the gauge freedom (\ref{13}) is
frozen- say by imposingg
$N=1, N^a = 0 $- the equations of motion following
from (\ref{16}) have a unique
solution. There is no futher degeneracy besides the one implied
 by (\ref{11}). [For
more on the general description of gauge symmetries, see \cite{r8} ].
Accordingly, there is complete agreement between the first geometric
definition of gauge symmetries and the dynamical definition based on the
Lagrangian.

The extra symmetries which the second geometrical definition allows
(elements
of the normalizer of $G$ which are not in the centralizer) fail to
be dynamical
gauge symmetries because, as we have already emphasized, they do not
 fulfill
(ii). Hence, they lead neither to a degeneracy of the Lagrangian nor
to an
indeterminacy of the Cauchy problem. Trying to ``gauge fix'' them is
unnecessary to get an unambiguous evolution and would  remove true
degrees of freedom. Even though these symmetries are not gauge symmetries,
they
do define, however, global symmetries as we now discuss.

\subsection{ Global symmetries}

The transformations (\ref{15}) with  $ \mu_a^b \not= 0$ induce the
following
changes of the dynamical variables ,
\begin{eqnarray}
\label{18}
 \delta N = 0 \nonumber \\
  \delta N^a = a^a_b N^b \\
    \delta g_{ab} = - a^c_b g_{cb} - a^c_b g_{ac} \nonumber
\end{eqnarray}
where $a^a_b$ is  a constant matrix belonging to the Lie algebra of the
automorphism group of $\cal G$. The transformations (\ref{18}) leave
the action
(\ref{16}) invariant provided $Tr \  a = 0$ and hence, define global
symmetries
when $Tr \  a = 0$.  By Noether theorem, they give non vanishing
 conserved
charges.

A crucial feature of  the transformations (\ref{18}) of the dynamical
variables is that they  are well defined even if the spacial sections are
compact. In that case, they cannot be induced by a globally defined
diffeomorphism\cite{r3}, but still, they make perfect sense.
Futhermore, they take exactly the same form irrespectively of the
topology. From the dynamical point of view, they should always be
regarded
as global symmetries since they cannot be localized in $x^0$ by a
choice of
the parameters $a^a_b$. In the type I case, the corresponding conserved
charges account for the integrability of the model \cite{r1,r2}.

\subsection{ Counting degrees of freedom- Covariant phase space}

The number of degrees of freedom of a gauge theory can be counted
directly from the Lagrangian, without having to go through the
Hamiltonian analysis. It is equal to (half) the dimension
of the covariant
phase space, defined to be the space  of solutions of the equations
 of motion
modulo the gauge transformations \cite{r9,r10,r11,r12}.

        Now, a solution of the equation of motion is completely
characterized by: (i) a choice of lapse $N(x^0)$ and
shift $N^a(x^0)$; and
(ii) initial data $g_{ab}( x^0 = 0)$, $
\frac{dg_{ab}}{dx^0}(x^0 = 0)$ at $x^0
= 0$ (say) fulfilling the constraint equation
 \begin{eqnarray}
\label{19}
G^{00} = \frac{ \sqrt{(N^2 -N^s N_s)}}{N^2}
    (R + K^i_i K^j_j - K^i_j  K^j_i) \\
     G^{0i} = \frac{2}{N} (- {{C^l}_l}^i K^j_j + {C_{lj}^i}
K^{lj}) +  \frac{N^i}{N^2} (R + K^i_i K^j_j - K_{ij} K^{ij}
)\nonumber \end{eqnarray}
The number of independent equations (\ref{19}) is equal
 to $4 - dim ( center
\cal G)$.
By a choice of gauge, on can take $N=1$ and $N^a = 0$.
The residual gauge
transformations acting on the initial data $g_{ab}$,
 $\frac{dg_{ab}}{dx^0}$
take the form
\begin{eqnarray}
\label{20}
  \delta g_{ab} = \mu \frac{dg_{ab}}{dx^0} + \mu^{\delta}
   ( C_{\delta a} g_{cb} + C_{\delta b}^c g_{ac})  \\
\delta \frac{dg_{ab}}{dx^0} = \mu [ -2 R_{ab} - \frac{1}{2}
\frac{dg_{cd}}{dx^0} g^{cd} \frac{dg_{ab}}{dx^0}
 +\frac{dg_{bl}}{dx^0} g^{kl} \frac{dg_{ak}}{dx^0}] + \mu^\delta
    ( C_{\delta a}^c \frac{dg_{cb}}{dx^0} + C_{\delta b }^c
\frac{dg_ac}{dx^0}) \nonumber
\end{eqnarray}
and involve effectively $4 - dim ( center \cal G)$ parameters
( recall that $
\mu^a = ( \mu^A , \mu^\delta )    $ where $\mu^A$ is in the center).
 Hence,
the dimension of the space of solutions of the equations of motion modulo
the gauge-transformation is equal to 12 (number of $g_{ab}$ and
$\frac{dg_{ab}}{dx^0}$, $N$ and  $N^a$ do not count since they can be
``gauged  away") minus 4 -  dim(center $\cal G$) ( number of independent
transformations (\ref{20}). The number of degreees of freedom is half
this count,
\begin{eqnarray}
\label{21}
 n  & = &\frac{1}{2} \{ 12 - 2 [ 4 - dim ( center \  \cal{G} ) ] \} \\
      & = & 2 + dim ( center \ \cal{ G} ) \nonumber
\end{eqnarray}

The count (\ref{21}) agrees with the count of \cite{r8},
 Chapter 3, where it is shown
that $n$ is equal to the number of dynamical variables
( here 10 = number of
$g_{\mu \nu}$) minus the number of times the gauge
transformations strike. A
gauge transformation strikes n times if the corresponding gauge
parameter together with its time derivatives  up to order n-1 appear
effectively in the gauge transformations]. Here, $\mu$ and $\mu^\alpha$
strike twice , but the gauge parameters $\mu^A$ in the center strike only
once. Hence, again,
\begin{eqnarray}
 n & = &10 - 2 \{1 + 3 - [ dim ( center\  {\cal G}) ] \} -  dim( center\
 {\cal G} )
       \\ & = & 2 + dim ( center\ \cal{ G})
\end{eqnarray}

The symplectic structure on the covariant phase space is determined by the
boundary term in the variation in the action. One finds here (on-shell)
   \begin{equation}
 \delta S = [ \Pi^{ab} \delta g_{ab} ]^{x^0_2}_{x^0_1}
\end{equation}
with
   \begin{equation} \Pi^{ab} = \sqrt{g} (g^{ab} K^i_i - K^{ab}) \nonumber
\end{equation}
Thus, the symplectic two-form on the covariant phase space is the two-
form induced by the canonical two-form
 \begin{equation}
 \delta \Pi^{ab}  \wedge \delta g_{ab}
\end{equation}
on the quotient of the surface (\ref{19}) by the transformations
(\ref{20}).
[ One easily verifies that the induced two-form on the surface (\ref{19})
has  a rank equal to $4 - dim (center \cal G)$, with zero eigenvectors
determined by (\ref{20}). It is thus regular on the quotient space.]

\section{ Hamiltonian analysis}

The reader familiar with the hamiltonian analysis of Bianchi models will
certainly have noticed that the above results, obtained from a spacetime
 covariant point of
view, are just the same as the Hamiltonian results
(see e.g. \cite{r2,r3}). The
$\Pi^{ab}$ are the momenta conjugate to $g_{ab}$,
 the constraints (\ref{19}) are
the Hamiltonian constraints and are first class.
The transformations (\ref{20})
are just the transformations which they generate.
The details may be found in the
literature and will not repeated here.
We can thus conclude not only that the various concepts of gauge
transformations do coincide, but also, that the covariant phase
space is
identical (as symplectic manifold) with the Hamiltonian reduced
 phase space.
This later point is of course not an accident since it is been
 established
 on general grounds
in \cite{r13}.

\section{ Coset spaces}

One can generalize Bianchi models by assuming local homogeneity instead
of global homogeneity. In the globally homogeneous case, the surface of
transitivity can be identified with the homogeneity group $G$.
In the locally
homogeneous case, one replaces the surfaces of transitivity $G$ by the
quotient space $G / K$ where $K$ is a discrete subgroup of $G$
acting on $G$
properly discontinously. The diffeomorphisms of $G$ that do not commute
with $K$ fail to  globally defined on the quotient space. Homogeneous
objects do remain well-defined, however. To study coset spaces, it is
convenient to lift everything from $G / K$ to $G$. One then applies the
techniques of the previous sections. One finally takes the quotient
 at the
very end. Since the gauge transformations are generated by the vector
fields $X_a$, which commute with $\xi_a$ and hence with $G$ and $K
\subset G $, they are well defined in the quotient.

\section{ Remarks on Bianchi models and Ashtekar variables}

Recently, Ashtekar has introduced new variables in terms of which the
constraints of general relativity take a polynomial form
 ( see e. g.  the
reviews [14,15]). In order to gain insight into those variables,
 various
authors have reformulated Bianchi models in terms of them \cite{r16,r17}.
An interesting question then arises: what is the most general form of the
Ashtekar variables compatible with homogeneity? We first address
 this question.

A SO(3)-connection $A_{(a)k}$ and a triad  $e_{(a)k}$  are homogeneous if
and only if the images  $h^*_R A_{(a)k}$ and $h^*_R e_{(a)k}$
 of $A_{(a)k}
$ and $e_{(a)k}$ differ from $A_{(a)k}    $ and $e_{(a)k}$
 by a local SO(3)-
transformation \cite{r18,r19,r20,r21,r22}
\begin{eqnarray}
\label{27}
 h^*_R  A_k = \psi A_k \psi_R^{-1} + \psi_R \partial_k \psi_R^{-1} \\
\label{27'} h^*_R e_k = \psi_R e_k
\end{eqnarray}
where $\psi_R$ is a spacerime-dependent element of
 SO(3) that depend also on
$R \in G$ . If we assume that  the triads are non-degenerate,
or degenerate
at most at isolated points, one infers from (\ref{27}),(\ref{27'})
 and  $h^*_{R_1 R_2} =
h^*_{R_1} h^*_{R_2}$ that
\begin{equation}
\label{28}
 \psi_{R_2 R_1}(P) = \psi_{R_2}(h_{R_1} P ) \psi_{R_1}(P)
\end{equation}
so that the difficulties pointed out in \cite{r23} and characteristic
of Abelian
gauge groups do not occur. One can then apply the general techniques
developed in \cite{r24,r25}.
Since the group $G$ acts simply transitively, one may define $\chi(P)$
as  \begin{equation}
\label{29}
 \chi(P) = \psi_R^{-1}(P_0)
\end{equation}
where $P_0$  is a fiducial reference point on the surfaces of
transitivity to which $P$ belongs
 ( for
instance, one can take it to be the intersection point of $\Sigma$
with the
curve $\gamma$ )  and   $R$ is the unique element of $G$ such that
$h_R(P_0) = P$.

Let us define $\tilde{A}$ and $\tilde{e}  $ by making the S0(3) gauge
transformation generated by $\chi$,
\begin{eqnarray}
\label{30}
 \tilde{A} = \chi A \chi^{-1} + \chi d\chi^{-1} \\
     \tilde{e} = \chi e \nonumber
\end{eqnarray}
There, one finds from (\ref{27}),(\ref{27'}),(\ref{28}) and
(\ref{29}) that $\tilde{A}$ and $\tilde{e}  $
are strictly invariant,
\begin{eqnarray}
\label{31}
 h^*_R \tilde{A} = \tilde{A}\\
    h^*_R \tilde{e} = \tilde{e}
\end{eqnarray}

There is thus no loss of generality  in assuming the Ashtekar
 variables to
be strictly homogrneous. This confirms by a geometrical method the
findings of \cite{r16,r17} . In the invariant frames   $\{ X_a \}$,
the components of $\tilde{A}_{(a)k}$ and $\tilde{e}_{(a)k}$ depend
only on
$x^{0}$ . The Einstein equations define a problem of classical
mechanics,
as in the metric formulation.

The definition of the gauge symmetries proceeds then as above.
These symmetries are
found to include besides the homogeneity preserving diffeomorphisms
of the metric
formulation , the local SO(3) internal transformations acting on the
internal indices. There is again complete
 agreement between the first geometrical definition and the dynamical
 definitions.

The analysis  can be extended to the discussion of
discrete symmetries (see \cite{Sc,Ma})  provided we assume the
connection to be an O(3)-connection rather that an SO(3)-connection .
 To illustrate the point, let us consider
Bianchi type I models,  $ds^2 = - N^2(x^0) (dx^0)^2 +
g_{ab}(x^0) dx^a dx^b$,
with the discrete symmetry $x^3 \mapsto -x^3$. Then  $g_{13} = g_{23} =
0$. In terms of the triad $e_{(a)k}(x^0)$ and the connection
$\omega_{(a)k}(x^0)$, the condition of invariance becomes
\begin{eqnarray}
\label{32}
 e'^{(a)}_k = {\chi^{(a)}}_{(b)} e^{(b)}_k \\
\omega'^{(a)}_k = {\chi^{(a)}}_{(b)} \omega'^{(b)}_k
  + \frac{1}{2} \varepsilon^{(abc)} \chi_{(bd)}
\partial_k {\chi^{-1}}_{(c)}^{(d)}
\end{eqnarray}
where $e'^{(a)}_k = ( e^{(a)}_1,e^{(a)}_2,-e^{(a)}_3)$ and
 $\omega'^{(a)}_k = ( \omega^{(a)}_1,\omega^{(a)}_2,-\omega^{(a)}_3)     $

Equation (\ref{32}) shows that $   {\chi^{(a)}}_{(b)}    $ has two
eigenvalues equal
to +1 and one equal to -1. Hence it does not belong to SO(3) but rather
to
O(3). In this case, one must extend (\ref{32}) in the following way:
\begin{eqnarray}
 e'^{(a)}_k = {\chi^{(a)}}_{(b)} e^{(b)}_k \\
\omega'^{(a)}_k = -{\chi^{(a)}}_{(b)} \omega'^{(b)}_k
  + \frac{1}{2} \varepsilon^{(abc)} \chi_{(bd)}
\partial_k {\chi^{-1}}_{(c)}^{(d)}
\end{eqnarray}
for elements ${\chi^{(a)}}_{(b)}$ of O(3) which are not
in the connected component to the identity. Indeed, $\omega'^{(a)}_k$
is a pseudo-vector for O(3) transformations.

 In an appropriate basis, the matrix ${\chi^{(a)}}_{(b)}$ reads

\begin{equation}
  {\chi^{(a)}}_{(b)}  = \left(  \begin{array}{ccc}
 1 & 0 & 0 \\ 0 & 1 & 0 \\ 0 & 0  & -1 \end{array} \right)
\end{equation}

and is constant. The invariance condition implies then
\begin{eqnarray}
  e^{(1)}_3 = e^{(2)}_3 = 0 ,  e^{(3)}_1 = e^{(3)}_2 = 0 \\
 \omega^{(1)}1 = \omega^{(1)}_2 = 0 ,  \omega^{(2)}_1 = \omega^{(2)}_2
 = 0
\end{eqnarray}

The variables remaining in the reduced variationnal principle for
Bianchi type
 $\underline {\overline I} $ cosmologies with discrete symmetry
$ x^3 \mapsto -x^3$ are thus:
\begin{itemize}
\item[(i)] $e^{(a)}_k, \omega^{(3)}_k \hskip 1 cm ( a = 1,2 ;
\  k = 1 , 2 )$,
\item[(ii)] $e^{(3)}_3,\omega^{(k)}_3 \hskip 1 cm ( a = 1,2) $,
\item[(iii)] $N$ and $N^k$ (which drops out).
\end{itemize}
 The gauge symmetries are (i) the redefinition of the time
variable; and (ii) the SO(2)-rotations in the (1)-(2) plane.

\section{Conclusion}

We have shown that  there exists a natural definition of homogeneity
preserving diffeomorphisms for Bianchi models such that none of the
difficulties found in  [3] arises. With that definition, the geometrical
 and
dynamical concepts of gauge symmetries coincide irrespectively  of the
spatial topology. We have also verified the isomorphism of the Lagrangian
covariant phase space with  the Hamiltonian reduced phase space. Finally,
we have discussed spatially homogeneous Ashtekar variables. We have
 illustrated how the
general  theory of invariant Yang-Mills connections apply. This theory
 is actually
 relevant in the discussion of all minisuperspace models formulated in
terms of Ashtekar variables, and not just the Bianchi cosmological models
 with a three-dimensional
simply transitive invariance group. In particular, one can generalize
 the considerations of \cite{ku}
section $\overline {\underline V}$ to the Ashtekar description.

\section{Acknowledgements}

We are grateful to Luca Bombelli for useful discussions.
This work has been supported an part by the FNRS(Belgium)
and the Commision of the European Communities.


\begin{thebibliography}{99}

\bibitem{r1}M.Henneaux,
``Sur la g\'eom\'etrodynamique avec les champs
spinoriels'', chapter V, PhD Thesis (Brussels 1980).
 \bibitem{r2}  M.Henneaux,Ann. Inst.
Henri Poincar\'e \underline{34}(1981)329; see also
M.Henneaux, Lett. Nuovo Cimento
\underline{38}(1983)609(especially page
612).
\bibitem{r3} A.Ashtekar and J.Samuel,
Class.Quant.Grav.\underline{8}(1991)2191.
\bibitem{r4} R.T.Jantzen,
Commun.Math.Phys.\underline{64}(1979)211.
\bibitem{r5} M.A.H.Mac Callum, in ``General Relativity: An
Einstein Centenary Survey'', S.W.Hawking and W.Israel eds,
Cambridge University Press (Cambridge: 1979).
\bibitem{r6} G.E.Sneddon, J.Phys.\underline{A9}(1975)229.
\bibitem{r7} A.H.Taub, Ann.Math.\underline{53}(1951)472.
\bibitem{r8} M.Henneaux and C.Teitelboim, ``Quantization of
Gauge Systems'', Princeton University Press(Princeton:
1992).
 \bibitem{r9} J.Kijowski and W.Szczyrba,
Comm.Math.Phys.\underline{46}(1976)183.
\bibitem{r10} E.Witten,
Nucl.Phys.\underline{B276}(1986)291.
\bibitem{r11} G.Zuckerman, Yale University Preprint(1986).
\bibitem{r12} A.Ashtekar, L.Bombelli and O.Reula, in
``Analysis, Geometry and Mechanics: 200 years after
Lagrange'', eds.M.Francaviglia and D.Holm, Noth Holland,
Amsterdam 1990.
 \bibitem{r13} G.Barnich, M.Henneaux and
C.Schomblond, Phys.Rev.\underline{D44}(1991)R939.
\bibitem{r14} A.Ashtekar, Lectures on Non-Perturbative
Canonical Gravity, World Scientific, Singapore (1991).
\bibitem{r15} C.Rovelli,
Class.Quant.Grav.\underline{8}(1991)1613.
\bibitem{r16} H.Kodama, Prog.Theor.Phys., 80(1988)1024.
\bibitem{r17} A.Ashtekar and J.Pullin,
Ann.Israel.Phys.Soc. 9(1990)65
\bibitem{r18} P.Cordero and
C.Teitelboim, Ann.Phys.(N.Y.) \underline{100}(1976)603;
R.Benguria,P.Cordero and C.Teitelboim,
Nucl.Phys.\underline{B122}(1977)61. \bibitem{r19}
V.N.Romano,A.S.Schwarz and Yn.S.Tyupkin, Nucl.Phys.
\underline{B130}(1977)209. \bibitem{r20} P.G.Bergmann and
E.J.Flaherty, J.Math.Phys\underline{19}(1978)212.
\bibitem{r21} A.Trautman, J.Phys.\underline{A13}(1980)L1.
\bibitem{r22} A.S.Schwarz,
Comm.Math.Phys.\underline{56}(1977)79.
\bibitem{r23} M.Henneaux,
J.Math.Phys.\underline{23}(1982)830.
\bibitem{r24} J.Harnad,S.Shnider and L.Vinet,
J.Math.Phys.\underline{20}(1979)931.
\bibitem{r25} P.Forg\'acs and N.S. Manton,
Comm.Math.Phys.\underline{56}(1977)79.
\bibitem{Sc} B.G.Schmidt,
Comm.Math.Phys.\underline{15}(1970)329. \bibitem{Ma}
M.A.H.Mac Callum, in Carg\`ese lectures in Physics volume
6, ed.E.Schatzman, Gordan and Beach (New York 1973).
\bibitem{ku} K.Kuch\v ar,
Phys.Rev.\underline{D4}(1971)955. \end{thebibliography}
\end{document}